\documentclass[%
 reprint,
superscriptaddress,
 amsmath,amssymb,
 aps,
]{revtex4-1}

\usepackage{graphicx}
\usepackage{dcolumn}
\usepackage{bm}

\begin{document}

\preprint{APS/123-QED}

\title{Plasmon Localization Assisted by Conformal Symmetry}

\author{Lizhen Lu}
\affiliation{%
State Key Laboratory of Electrical Insulation and Power Equipment, School of Electrical Engineering,
Xi'an Jiaotong University, Xi'an 710049, China
}%
\affiliation{
 The Blackett Laboratory, Department of Physics, Imperial College London, London SW7 2AZ, United Kingdom
}%
\author{Emanuele Galiffi}
 \email{emanuele.galiffi12@imperial.ac.uk}
\affiliation{
 The Blackett Laboratory, Department of Physics, Imperial College London, London SW7 2AZ, United Kingdom
}%
 
\author{Kun Ding}
 \email{k.ding@imperial.ac.uk}
\affiliation{
 The Blackett Laboratory, Department of Physics, Imperial College London, London SW7 2AZ, United Kingdom
}%

\author{Tianyu Dong}
\affiliation{%
	State Key Laboratory of Electrical Insulation and Power Equipment, School of Electrical Engineering,
	Xi'an Jiaotong University, Xi'an 710049, China
}%
\author{Xikui Ma}
\affiliation{%
	State Key Laboratory of Electrical Insulation and Power Equipment, School of Electrical Engineering,
	Xi'an Jiaotong University, Xi'an 710049, China
}%
\author{John Pendry}
\affiliation{
 The Blackett Laboratory, Department of Physics, Imperial College London, London SW7 2AZ, United Kingdom
}%

\date{\today}

\begin{abstract}
Plasmonic systems have attracted remarkable interest due to their application to the subwavelength confinement of light and the associated enhancement of light-matter interactions. However, this requires light to dwell at a given spatial location over timescales longer than the coupling rate to any relevant loss mechanism. Here we develop a general strategy for the design of stopped-light plasmonic metasurfaces, by taking advantage of the conformal symmetry which underpins near-field optics. By means of the analytical technique of transformation optics, we propose a class of plasmonic gratings which is able to achieve ultra-slow group velocities, effectively freezing surface plasmon polaritons in space over their whole lifetime. Our method can be universally applied to the localization of polaritons in metallic systems, as well as in highly doped semiconductors and even two-dimensional conductive and polar materials, and may find potential applications in nano-focusing, nano-imaging, spectroscopy and light-harvesting.
\end{abstract}


\maketitle
\section{Introduction}

Symmetry is one of the most elegant and useful concepts in physics, constituting a powerful tool to simplify complex problems, as well as understanding their underlying physical mechanisms~\cite{dresselhaus2008applications,sakoda2004optical}. A specially useful tool for the harnessing of symmetry in electromagnetism is transformation optics (TO)~\cite{ward1996refraction,pendry2006controlling,leonhardt2006optical}. Originally conceived as an insightful strategy for the design and control of electromagnetic fields, TO has more recently played a fundamental role in the modeling of plasmonic systems, providing valuable insight into their electromagnetic response. In particular the technique has been applied to the study of a class of gratings on metal films generated by conformal transformations of a slab of metal whose Bloch eigenmodes are dictated by its translational symmetry~\cite{huidobro2010transformation,pendry2012transformation,kraft2015designing,kraft2014transformation}. At the center of the Brillouin zone (BZ), $k=0$, the frequencies of the modes are always those of the original slab, no matter what the parameters of the transformation provided that it is conformal. Although the translational symmetry of the slab is hidden by the transformation its presence is felt at the center of the BZ. Away from the BZ center, at the boundary, a gap opens: the stronger the modulations of the grating, the wider the gap. Since the band remains pinned at the center, this has the effect of compressing the band, reducing the group velocity and hence stopping the light. Our strategy is to optimize localization by tuning the parameters of the conformal transformation. 

The phenomenon of slow light has been a fundamental breakthrough in optics, sustained, on the one hand, by the advances in the manipulation of material dispersion via nonlinear schemes~\cite{hau1999light,boyd2009controlling}, and on the other by the advent of photonic crystals and metamaterials enabled by recent progress in nano-fabrication~\cite{settle2007flatband,baba2008slow,yu2014flat,kildishev2013planar,shaltout2019spatiotemporal}. Tailoring of dispersion has thus enabled realization of flat bands, associated with extremely low group velocities~\cite{tsakmakidis2007trapped,tsakmakidis2014completely,tsakmakidis2017ultraslow,gan2008ultrawide,hao2019increasing,kubo2007low,minkov2015wide,hao2010novel}. Surface plasmon polaritons (SPPs) are well-known for slowing down the propagation of light while confining it to subwavelength volumes, acting as efficient platforms for the enhancement of light-matter interactions and optical nonlinearities, with applications ranging from sensing to quantum optics, optical communications, data storage and energy harvesting~\cite{woessner2015highly,chubchev2018highly,kauranen2012nonlinear,luo2013harvesting,chon2007spectral,kongsuwan2019quantum,mansuripur2009plasmonic,luo2012broadband}. However, conventional plasmon localization requires meticulous and time-consuming design of surface structures to minimize dissipative and radiative losses. Therefore, establishing an insightful methodology to achieve flat SPP bands in metallic metasurfaces is a challenge worth pursuing. 

In this work we exploit conformal symmetry as a new insightful guide for the design of metasurfaces with extremely flat bands, enabling the localization of SPPs over timescales much longer than their lifetime, whilst minimizing radiative losses. This paper is organized as follows: in Sec.~\ref{sec:Realization of flat bands by conformal symmetry} we show how conformal mapping can be used to generate a series of metasurfaces starting from a translationally invariant slab, and how the resulting conformal symmetry can be exploited to leverage the plasmonic band structure in order to realize extremely flat bands. We then develop in Sec.~\ref{sec:Analytical study of plasmon localization} an intuitive analytic framework to predict the temporal dynamics of SPPs based on the dispersion bands thus calculated. In Sec.~\ref{sec:Numerical Demonstration of plasmon localization} we corroborate our findings against frequency-domain and time-domain simulations of realistic physical setups, demonstrating the ability of conformal metasurfaces to effectively freeze light in the nano-scale over timescales exceeding the lifetime of plasmons, while introducing negligible radiative loss, thus yielding a class of plasmonic surfaces with promising potential for the enhancement of light-matter interactions. Finally, we present our conclusions and final remarks in Sec.~\ref{sec:conclusions}.

\section{Realization of flat bands by conformal symmetry}
\label{sec:Realization of flat bands by conformal symmetry}
In a nano-plasmonic system, the electric field $\mathbf{E}$ is determined primarily by the electrostatic potential $\Phi$, since its fast spatial oscillations dominate over the retarded contribution. Thus, the Laplace equation $\nabla^2 \Phi = 0$ governs the subwavelength dynamics of SPPs. If the surface is periodic along the $y$-axis, a complete solution may be written in terms of Bloch modes $\Phi_{n,k_y}(x,y) = e^{i k_y y}\phi_{n,k_y}(x,y)$, where the eigenfunctions $\phi_{n,k_y}(x,y) = \phi_{n,k_y}(x,y+a)$, and $k_y$ is the Bloch wavevector of the plasmon. Thus, at $k_y=0$, the periodic eigenfunctions $\phi_{n,k_y=0}(x,y)$ are themselves solutions of the Laplace equation.

\begin{figure}[ht]
\centering
\includegraphics[width=0.9\columnwidth]{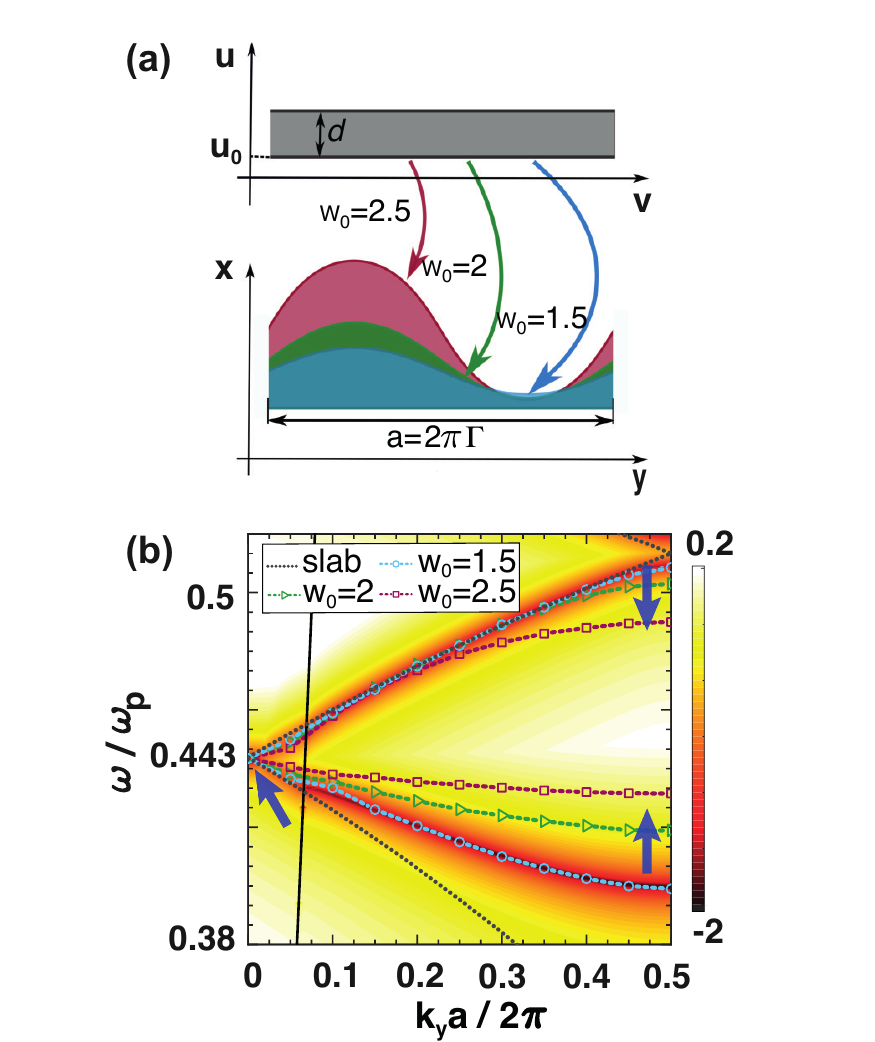}
\caption{(a) Schematic illustration of a class of conformal plasmonic metasurfaces: the gratings in the physical $x-y$ frame are obtained from the flat slab in the virtual $u-v$ frame via the conformal transformation (Eq.~\ref{eq:mapping}) with different modulation depths $\text{w}_0 = 1.5$ (blue), $\text  w_0 = 2.0$ (green) and $\text  w_0 = 2.5$ (red), generating gratings with modulation depths of $12.5$, $19.9$ and $37$ nm respectively. The period $a = 2\pi\Gamma \approx 94.2$ nm, and we assume a Drude metal with plasma frequency $\omega_{\text{p}} = 2$ eV and loss rate $\gamma_e = 2$ meV. (b) Colormap of $ \max{ \{1/\text{eig}|\mathbf{R}(\omega ,k_y)| \} }$ in log-10 scale for the grating with $\Gamma=15$ nm, $d=0.5$, $u_0=1.0$, and $\text  w_0=1.5$. Eigenspectra calculated numerically for gratings with $\text  w_0=1.5$, $\text  w_0=2$, $\text  w_0=2.5$, are shown as blue, green and red lines, respectively, showing the flattening of the band with increasing modulation strength $\text  w_0$. The dotted line shows the dispersion of the flat slab (thickness $7.5$ nm) folded into the first BZ, which coincides with the conformal gratings at $k_y=0$ due to the conformal symmetry. The solid black line corresponds to the light line.}
\label{fig:1}
\end{figure}

The Laplace equation has been known to be symmetric under a conformal coordinate transformation since the early days of electrostatics, a strategy often used to calculate analytically the electrostatic field for complicated geometries by transforming the known analytic solutions of simpler structures via conformal maps~\cite{schinzinger2012conformal}. However, the band-folding of the dispersion relation in periodic systems implies that these structures feature an infinite set of points in reciprocal space, located at the centre of the BZ ($k_y=0$) where conformal symmetry is obeyed, and yet the associated modes have finite resonant frequencies. This symmetry may be leveraged for the purpose of designing a whole class of slow-light structures, as explained below. Consider the system shown in Fig~\ref{fig:1} (a): a series of gratings in the (physical) $x-y$ frame can be mathematically generated from a single slab positioned at $u=u_0$, in the (virtual) $u-v$ plane, with constant thickness $d$, via the following conformal transformation:
\begin{equation}
z=\Gamma\ln{\bigg[\frac{1}{e^\text{w}-\rm{i} \text  w_0}+\rm{i}y_0\bigg]}
\label{eq:mapping}
\end{equation}
with $z=x+\rm{i}y$, and $\text{w}=u+\rm{i}v$. The scaling factor $\Gamma$ determines the period $a=2\pi\Gamma$ of the gratings, and $\text  w_0$ is the modulation depth~\cite{kraft2015designing}. Here, we choose $y_0=\text  w_0/[e^{2(u_0+d)}-\text  w_0^2]$, which ensures that the surface at $u=u_0+d$ remains flat upon the mapping. 

The slab and the gratings are assumed to consist of a conductive material described by a Drude model, with plasma frequency $2$ eV, which is a typical value for transparent conductive dioxides. In order to improve the visibility of the bands, in this section we choose a small value for the electron scattering rate $\gamma_e = 2$ meV, which, however does not affect the validity of our argument. The optical response is obtained via semi-analytical solutions calculated by transforming the full set of Maxwell's equations via transformation optics, and finding the poles of the reflection coefficient, as detailed in \cite{pendry2019computing,PhysRevB.100.115412}. In addition, we verify our semi-analytical calculations by plotting the corresponding spectra obtained via finite-element simulations performed with COMSOL Multiphysics.

It is worth emphasizing that the whole class of gratings in physical space spawns from the single slab in the virtual space, and the value of $\text{w}_0/\mathrm{e}^{u_0}$ determines the modulation depth of the grating. Hence, our conformal symmetry argument ensures that, for any value of $\text  w_0$, the resulting structures share their resonances with the virtual slab at $k_{y}=0$, \textit{i.e.} at the center of the BZ. Conversely, at the second high-symmetry point $k_y=\pi/a$ a band-gap opens, whose extent increases as the modulation-depth parameter $\text  w_0$ sweeps the range from $0$ to $e^{u_0}$, where the transformation features a geometrical singularity. Hence, the plasmonic spectrum of this system is symmetry-protected against any changes in $\text  w_0$ at $k_y=0$, but not away from it. This is illustrated in Fig.~\ref{fig:1}(b): the contour plot shows the band structure of a plasmonic grating generated with a modulation amplitude $\text w_0=1.5$, where the blue open dots depict the same bands calculated using COMSOL, thereby demonstrating excellent agreement. The green and red lines show the spectra for gratings with higher modulation strength $\text  w_0 = 2.0$ and $2.5$ respectively. Note that at the center of the BZ all gratings, regardless of their modulation strength $\text  w_0$, share the same resonance frequency, as a result of the conformal symmetry which relates them. This is determined solely by the analytic dispersion of the flat slab in the virtual $u-v$ frame~\cite{maier2007plasmonics}
\begin{align}
    e^{|k_y|d}=\pm\frac{\epsilon_m(\omega)-\epsilon_d}{\epsilon_m(\omega)+\epsilon_d},
\end{align}
which is shown in gray dotted lines. Here $\epsilon_m(\omega)$ and $\epsilon_d$ are the permittivities of the metal and of the surrounding dielectric medium, respectively. Hence, it is clear that the increase in modulation strength $\text  w_0$ of these conformal gratings acts as a lever for the plasmon band, yielding an elegant and powerful strategy to realize extremely flat bands, which are a signature of plasmon localization.

\section{Analytical study of plasmon localization}
\label{sec:Analytical study of plasmon localization}
\begin{figure}[t!]
\centering
\includegraphics[width=\columnwidth]{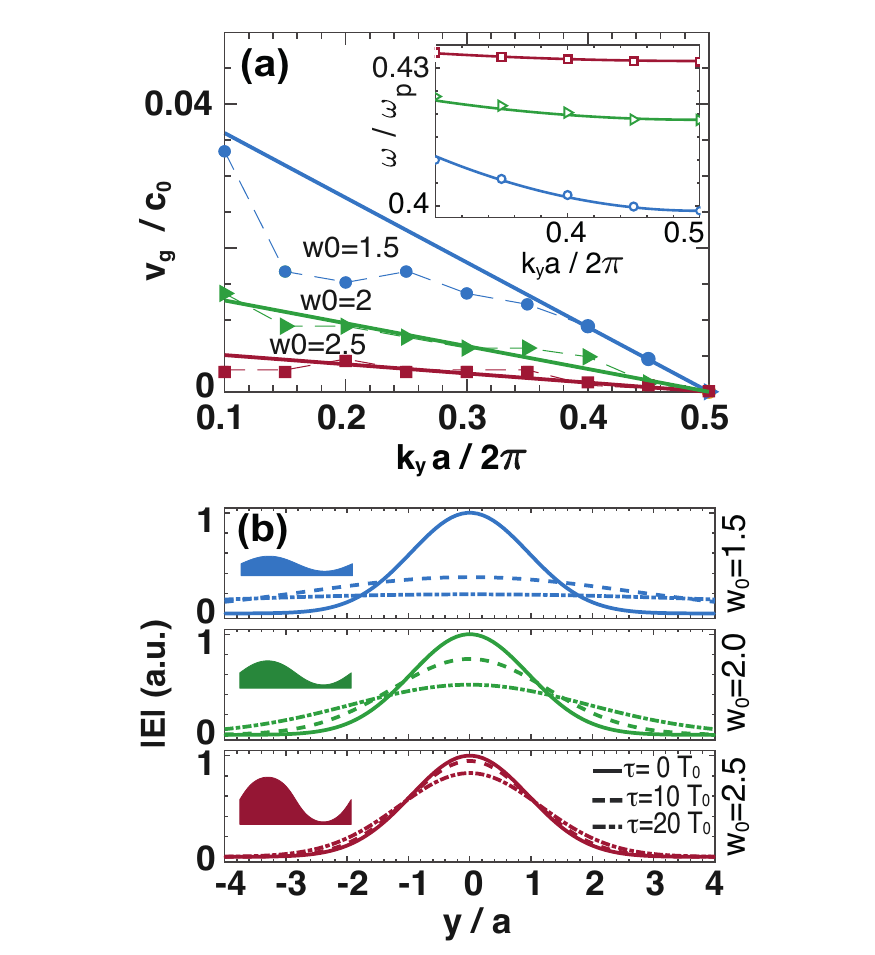}
\caption{(a) Group velocity $v_g/c_0$ of the plasmon bands for the three gratings considered, obtained from our COMSOL simulations (dots) and fitted to first order near the band gap (lines). The inset shows the fit of the band near the gap. \iffalse (b)  corresponding to gratings with modulation depths $\text  w_0 = 1.5$ (are shown by blue filled circles), $\text  w_0 = 2.0$ (green filled triangles) and $\text  w_0 = 2.5$ (red filled squares), respectively. The inset shows a blow up of the bands obtained from COMSOL simulations near the BZ edge (open dots), together with quadratic fittings the extracted  $v_g/c_0$ from the quadratic fitting functions (shown in the inset) for gratings with $\text  w_0=1.5$, $\text  w_0=2$, $\text  w_0=2.5$, respectively. Inset: Quadratic fitting curves (solid lines) to COMSOL simulated mode frequencies (open dots)(solid lines). \fi (b) Snapshots of the spreading of a gaussian pulse with initial full width at half maximum (FWHM) of $150 \rm{nm}$ at times $\tau=0 T_0$  (solid line), $\tau=10 T_0$ (dashed line) and $\tau=20 T_0$ (dashed-dotted line), obtained from the respective quadratic dispersion fits for the increasingly strong gratings (top to bottom), at their respective band-edge frequencies, where $T_0$ corresponds to the oscillation period for each grating.}
\label{fig:2}
\end{figure}

We now demonstrate the potential of this concept for slowing down light in the nano-scale. We focus on the lower frequency band in Fig.~\ref{fig:1}(b). the realization of plasmon localization entails two crucial features: (1) the suppression of the group velocity of SPPs and (2) the preservation of a SPP pulse over time, without significant spreading of the original wavepacket. The first aspect is determined by the slope of the SPP band, whereas the latter is governed by the second derivative of the dispersion curve near the edge of the BZ ($k_y\simeq \pi/a$). Hence, in order to formulate a concise whilst insightful description of the physics at play, it is instructive to perform a quadratic expansion: 
\begin{equation}
     \omega(\Tilde{k}_y)_i \simeq \beta_i(\Tilde{k}_y-\Tilde{k}_v)^2+c_i,\label{eq:fitting}
 \end{equation}
of the band of interest close to the band edge, which is shown in the inset of Fig.~\ref{fig:2}(a) from $\Tilde{k}_y=0.3$ to $\Tilde{k}_y=0.5$, where $\Tilde{k}_y = k_y/(2\pi/a)$ and $\Tilde{k}_v = 0.5$ are rescaled wavevectors, the latter corresponding to the BZ edge. The suffix $i\in\{1,2,3\}$, refers to the modulation depths $\text  w_0=1.5$, $\text  w_0=2.0$, and $\text  w_0=2.5$ respectively. Note that the vertex of the parabola is located at $\Tilde{k}_v=0.5$. Hence, we can easily obtain the quadratic coefficients $\beta_1 \approx 0.592$ eV, $\beta_2 \approx 0.208$ eV and $\beta_3 \approx 0.084$ eV and the band-edge frequencies $c_1 \approx 0.798$ eV, $c_2 \approx 0.8375$ eV and $c_3 \approx 0.863$ eV. 

By differentiating the quadratic fitting functions (Fig.~\ref{fig:2}(a), inset) we can calculate the group velocity of plasmons with frequencies close to the band-gap frequencies $\omega_i$, which are shown as the solid lines in Fig.~\ref{fig:2}(a). The filled dots are obtained via numerical differential of the simulated dispersion shown in Fig.~\ref{fig:1}(b). The approximation is accurate over a broader region of reciprocal space for the two stronger grating cases ($\text  w_0=2$, $2.5$), whereas for the weak grating ($\text  w_0=1.5$) the exact dispersion departs from the quadratic expansion sooner, due to the stronger contribution of higher order terms in the dispersion curve. It is apparent how the group velocity is consistently reduced over a wider range of plasmon momenta, as the modulation depth increases. 

Having parametrized our dispersion bands, we now investigate the temporal evolution of a surface plasmon with an initial gaussian profile. Its reciprocal-space envelope reads $E(k_y)=e^{-\alpha (k_y-k_v)^2}$, where $\alpha$ determines the pulse width. The spatial width of the pulse is proportional to $\alpha$, meaning that larger $\alpha$ corresponds to broader spatial distributions of the pulse. Our quadratic expansion of the lower band of our three conformal gratings allows us to calculate the curvature of the bands near $\Tilde{k}_y = 0.5$, which determines the variation of group velocities, and hence the spreading rate of the pulse upon propagation along the surface. We can thus calculate analytically the temporal evolution of a gaussian pulse for each grating, which is given by \cite{jackson1999classical}: \begin{equation}
f(y,t)= \sqrt{\frac{\pi}{(\alpha+\text{i}\beta_i t)^2}}\text{exp}\left[-\frac{y^2(\alpha-\text{i}\beta_i t)}{4(\alpha^2+(\beta_i t)^2)}+\text{i} k_v y-\text{i} \omega_v \emph t\right],
\label{eq:pulse}
\end{equation}
where $\omega_v$ is the central frequency of the pulse, and we substituted the dispersion relation (Eq. \ref{eq:fitting}). From Eq.~\ref{eq:pulse}, the temporal evolution of the width and peak amplitude of the pulse can be evaluated as: \begin{equation}
   \frac{\Delta y_0(t=t_1)}{\Delta y_0(t=0)}=\sqrt{1+\left(\frac{\beta_i t_1}{\alpha}\right)^2},
\end{equation} and \begin{equation}
       \frac{\text{max}[|\mathbf{E}|(t=t_1)]}{\text{max}[|\mathbf{E}|(t=0)]}=\frac{1}{\sqrt{1+(\beta_i t_1/\alpha)^2}},
\end{equation} respectively. Our model allows us to determine the SPP localization timescale as the ratio $\alpha/\beta_i$ between the initial width of the pulse and the curvature of the plasmon bands. More specifically, a larger $\alpha/\beta_i$ ratio improves localization by reducing the slope of the group velocity ($\beta_i$) near the BZ edge.

Figure~\ref{fig:2}(b) shows our theoretical predictions for plasmon localization in the three metasurfaces considered. Here, the initial width in real space is defined by the full width at half minimum (FWHM) corresponding to $150$ nm. A weak grating (top panel) is not able to localize the plasmon, which soon spreads out. However, as the modulation strength is increased to $\text w_0 = 2.0$ (middle panel), the spreading of the initial SPP pulse over time slows down, preserving its shape over approximately 10 cycles. Finally, the SPP pulse is effectively unchanged over up to 20 cycles in the strongest modulation case $\text{w}_0=2.5$ (bottom panel). The above theoretical predictions indicate that light can dwell in a given region of space for much longer times if the surface is structured as a conformal grating. 

\section{Numerical Demonstration of plasmon localization}
\label{sec:Numerical Demonstration of plasmon localization}

In order to accurately evaluate the ability of conformal gratings to spatially localize surface plasmons and validate the above theoretical analysis, we now show the results of finite-element frequency-domain simulations performed in COMSOL Multiphysics for a finite system. The metasurface consists of 13 periods, and a dipole oriented along the x-axis is placed at its center, 100 nm above the bottom surface of the structures. The frequency of the source is tuned to match the band-edge frequency of the respective metasurfaces (see Fig.~\ref{fig:1}(b)), where the group velocities approach zero.

\begin{figure}[t]
\centering
\includegraphics[width=\columnwidth]{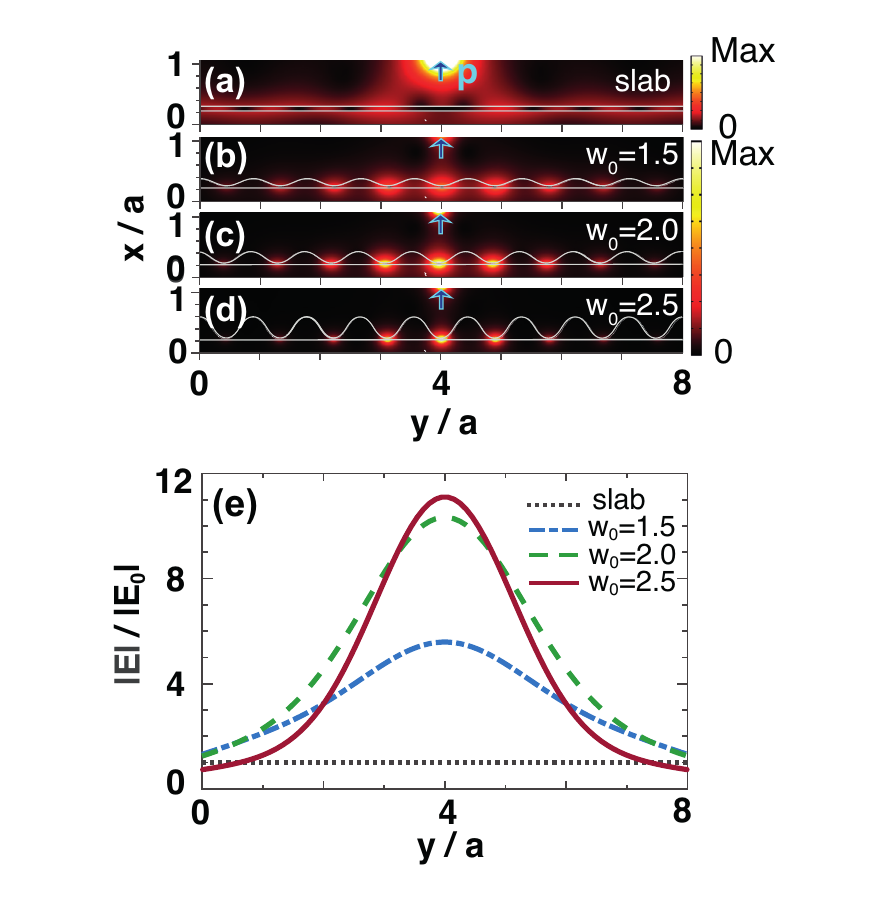}
\caption{Electric field distributions on the plasmonic slab (a) and metasurfaces (b)-(d) with different modulation depths $\text  w_0$ under electric dipole excitation (blue arrows) at their respective band-edge frequencies (as shown in Fig.~\ref{fig:2}(a)). (e) The envelope of the electric field distribution, obtained by interpolating the maximum electric field amplitude in each period of the metasurface, demonstrates the tighter localization achieved by metasurfaces with larger modulation depth $\text w_0$ (blue dot-dashed, green dashed and red continuous lines correspond to $\text w_0=1.5$, 2.0, and 2.5 respectively). In the flat slab case (black dashed line) the plasmon is completely delocalized.
}
\label{fig:3}
\end{figure}

\begin{figure*}[htbp]
\centering
\includegraphics[width=\linewidth]{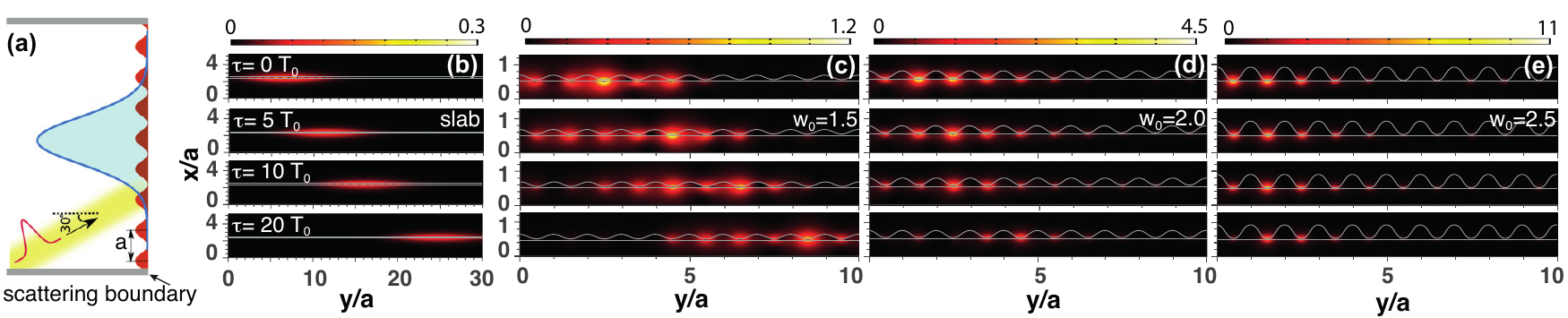}
\caption{(a) Schematic configuration used for testing the temporal behavior of surface plasmons on our conformal structures. (b)-(e) FETD simulations of the time evolution of light pulses injected from the left boundary towards the slab (b) and metasurfaces (c)-(e) at their respective band-edge frequencies. Increasingly strong conformal gratings stop the propagation of the pulse closer to the excitation region.}
\label{fig:4}
\end{figure*}

Figure~\ref{fig:3} shows the electric field distribution induced on a flat slab (a) and on our three increasingly strong gratings (b)-(d). In the slab case, the field distribution is completely delocalized. In fact, once excited by the near-field to the dipole, the SPP can freely travel away from it. Conversely, once a conformal metasurface is present, localization takes place: the electromagnetic energy of the SPP can no longer propagate, and it is effectively frozen in close proximity of the dipole. Plasmon localization can be further visualized  by plotting the envelope of the electric field amplitude (Fig. \ref{fig:3}(e)), calculated by interpolating the maximum amplitude between the different periods of the metasurface.

The narrower spatial width of the envelope for stronger modulation depths can be understood in real space as a result of the barriers formed by the metasurface, which dramatically slow down the SPP. However, a quantitative judgement of the optimal performance of conformal metasurfaces for localization comes from our description in reciprocal space, as the stronger Bragg scattering flattens the band across most of the BZ. Notably, our symmetry argument empowers us with deep insight into the scattering properties of these metasurfaces. Furthermore, by coupling the plasmon to large momenta, the eigenmodes of this system are extremely localized near the valley points of the structure, a feature previously studied in the context of singular surfaces \cite{pendry2017compacted,galiffi2019singular,yang2018transformation}.

Hence, the advantage of conformal metasurfaces for the enhancement of light-matter interactions is twofold: on the one hand the long dwelling of the plasmon near the excitation region opens interesting opportunities for, \textit{e.g.} the observation of coherent field dynamics between an emitter and the plasmonic surface. On the other hand, much stronger field enhancement takes place at the hot spots corresponding to the grating valleys, which can be strongly beneficial if emitters can be concentrated in these regions. Overall, our conformal strategy naturally reveals the non-trivial class of grating structures which, by exploiting the symmetry point at $k_y=0$, realize the most efficient route to localization.

In order to illustrate the temporal behaviour of plasmons in our conformal structures within a realistic experimental implementation with pulsed illumination, it is instructive to implement transient numerical simulations, carried out by means of a finite element time domain (FETD) method. As shown in Fig.~\ref{fig:4}(a), a Gaussian pulse is chosen as the excitation source, which can be depicted by the electric field:
\begin{equation}
    \mathbf{E}= e^{-(t-t_0)^2/(\Delta t)^2}e^{- i\omega_v t+ i\mathbf{k_0}\cdot\mathbf{r}},
\end{equation}
where the carrier frequency $\omega_v$ is chosen to match the band-edge frequencies of the respective metasurfaces as in the previous sections, the pulse duration $\Delta t=30$ fs, such that the frequency content of the pulse spans a range $\Delta\omega\approx 0.1$ eV, thus covering the entire lower band in the first BZ for all metasurfaces of interest here, and a time delay $t_0=80$ fs is used in the simulations. Due to the proximity of the injection port to the surface, the pulse contains evanescent components which can couple both to the metasurfaces and to the flat slab. In our simulations the plasmon excitation process lasts $\approx 155$ fs, after which we investigate its propagation along the surface.

Panels (b)-(e) in Fig.~\ref{fig:4} show the amplitude of the electric field at different times, corresponding to $0$, $5$, $10$ and $20$ oscillation periods $T_0$ (top-to-bottom rows), for a flat slab (column b) and for our three different metasurfaces (columns c-e). The starting point is chosen at 155 fs, when there is no excitation source. The velocity of the excited SPPs can be identified by following their center of energy\cite{tsakmakidis2014completely}:
\begin{equation}
    \mathbf{r} (t) = \frac{\int \mathbf{r}\cdot U(\mathbf{r}, t)\text{d}\mathbf{r}}{\int U(\mathbf{r}, t)\text{d}\mathbf{r}},
\end{equation}{}
where $U(\mathbf r,t)$ is the electromagnetic energy density. For the slab case, the excited plasmon travels at a speed $c_Y \approx c/22.5$, consistently with the analytical results derived from the well-known slab dispersion.
\begin{figure}[b!]
\centering
\includegraphics[width=\columnwidth]{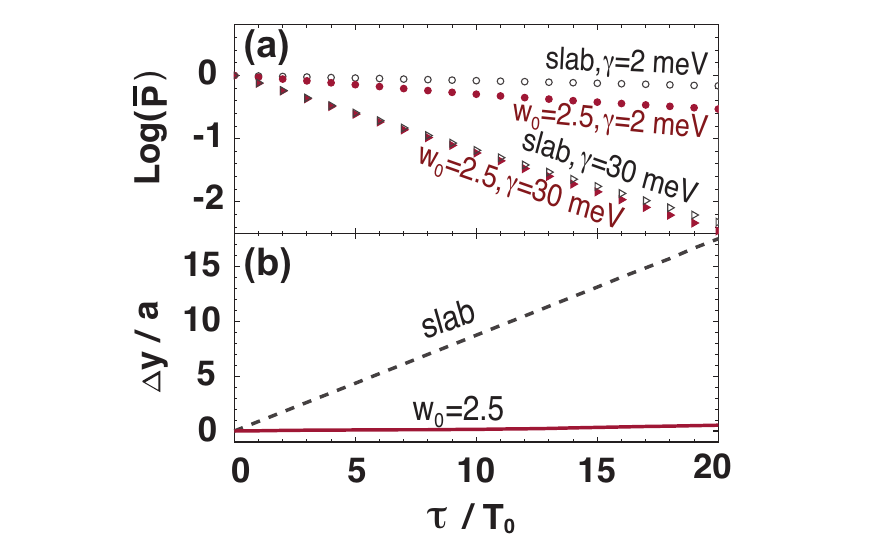}
\caption{(a) Variation of the total \emph{plasmon} energy over time for the excited SPP pulse on the $w_0 = 2.5$ metasurface (filled red markers) and on a flat slab (open black markers) assuming dissipative loss rates $\gamma_e=2$ meV (circles) and $30$ meV (triangles). The energy in the two cases is normalized to the respective total plasmon energies coupled to the surface immediately after the transient time. (b) The displacement of the center of energy of a SPP on a metasurface with $\text  w_0=2.5$ (red solid line) over time is negligible compared to the case of a flat slab, demonstrating that these structures can freeze the plasmon at the region illuminated by the pulse throughout its lifetime $\tau$, which is effectively only limited by the dissipative losses.}
\label{fig:5}
\end{figure}

Conversely, in the grating cases, the SPP pulse becomes increasingly slow (note the different y-axis scale), with center-of-mass velocities $c_Y \approx c/64.7$, $c/192.9$, and $c/1089.7$ for modulation strengths $\text  w_0 = 1.5$, $2.0$, and $2.5$ respectively, shown in Figs.~\ref{fig:4}(c)-(e) respectively. In particular, for the strongest grating, the pulse effectively appears to be frozen at the location where it is excited over more than 20 oscillation periods. In agreement with the frequency-domain simulations in Fig.~\ref{fig:3}, the SPPs in the three gratings also show tighter spatial confinement for stronger modulation strengths, as a consequence of the larger plasmon momenta achieved, which can concentrate the weight of the plasmon near the grating valleys. It is worth remarking that two competing mechanisms contribute to the finite velocity of the SPP pulse along the grating: firstly, part of the spectral content of the pulse entails frequencies which can propagate along the structure, even though their velocity is extremely low; this contribution has a dispersive nature, due to the different group and phase velocities sampled by the different pulse components. Secondly, the finite losses effectively cause a finite slope of the bands at the band-gap, so that even the carrier frequency of the pulse, which is tuned to match the band-gap frequency, is effectively able to propagate with finite group velocity. This latter effect, however, can be expected to have a minor impact.

Once experimental losses are taken into account, the plasmon lifetime $\tau = 1/\gamma_{tot}$, where $\gamma_{tot} \approx \gamma_{rad} + \gamma_{e}$, $\gamma_{rad}$ and $\gamma_e$ being the rate of radiative and dissipative decay respectively, sets the ultimate threshold for the longest meaningful localization time, since the plasmon would decay due to loss before delocalizing~\cite{khurgin2015deal,fukui1979lifetimes}. In Fig.~\ref{fig:5} we demonstrate that conformal gratings can localize SPPs over timescales which surpass by far their lifetime, whilst introducing negligible radiative losses. The top panel shows the decay of the plasmon energy over time for the slab (black) and the strongest grating considered ($\text  w_0=2.5$, red), assuming low (circles), as well as realistic (triangles) dissipative loss rates, corresponding to $\gamma_e=2$ meV and $\gamma_e= 30$ meV respectively. Panel~(b) shows, concurrently, the displacement of the center of energy for the slab (black dashed line) and grating (red continuous line), calculated based on our transient simulations. By observing the top and bottom plots simultaneously, it is evident that SPPs on a conformal grating are effectively frozen in space over a timescale much longer than their lifetime, whereas plasmons on a slab are free to drift away from the excitation point while their intensity is still high. 

In the low-loss case, the discrepancy between the decay rate of the slab and the grating allows us to estimate the the radiative losses suffered by our conformal metasurface as $\gamma_{rad}\approx 6.6$ meV. This radiative loss channel, \textit{i.e.} the scattering of SPPs into free-space radiation, only affects a corrugated surface, and it can be safely assumed to be uncoupled from dissipative mechanisms, so that increasing $\gamma_e$ will enhance the resistive loss, while having negligible effects on the radiative ones. In a typical realistic system with dissipative loss rate $\gamma_e = 30$ meV (Fig.~\ref{fig:5}(a), triangles) the total energy decreases by one order of magnitude within $\approx 8$ oscillation periods. We thus note that realistic dissipative loss rates strongly dominate over the radiative loss suffered by our conformal metasurface, indicating that the geometry of these structure offers optimal performance for localization while minimizing coupling of the SPP to the radiation continuum. Within a decay time $\tau_{tot}$, the position of the center of energy in the strong grating case only shifts by $\approx 11$ nm, which is only $11.7\%$ of the lattice length. This is in sharp contrast with the $\approx 7$ periods, which correspond to $\approx 660$ nm, travelled by the SPP along the flat slab within the same temporal window, thus demonstrating the power of conformal symmetry as a novel strategy for the design of slow-light plasmonic metasurfaces.

\section{Conclusions}
\label{sec:conclusions}

In this paper, we proposed a novel, general strategy to design plasmonic metasurfaces which are able to localize surface plasmons, by exploiting the conformal symmetry which underpins the near-field of a metallic nano-structure. We demonstrated how, by leveraging symmetry, these metasurfaces can feature extremely flat plasmon bands, achieving group velocities which are orders of magnitude lower than the speed of light in vacuum. We thus presented and thoroughly studied analytically and numerically a class of conformal gratings which can localize plasmons over timescales significantly longer than their lifetime, thus opening an interesting new avenue for symmetry-inspired design of stopped-light metasurfaces, which are able to combine strong light-matter interactions with the larger scattering cross-sections of an extended structure, compared to localized scatterers such as nano-particles. 

The present plasmon-localization concept can be applied using current nano-fabrication technology to any plasmonic material ranging from metals to semiconductors, as well as two-dimensional platforms such as graphene, as long as they are periodically structured on the subwavelength scale. In particular, by taking advantage of current low-loss plasmonic materials \cite{west2010searching,naik2013alternative}, such as transparent conducting oxides and doped semiconductors, our design strategy provides a powerful tool for realization of extended nano-devices, enabling a wide range of potential applications for enhanced light-matter interactions, nano-sensing, and energy-harvesting. Finally, the general strategy demonstrated here could be exploited for the localization of other surface waves on structured thin films, such as phonon polaritons in polar materials such as hexagonal Boron-Nitride, and our symmetry-based design concept could be extended to other wave phenomena, such as acoustic, elastic, seismic and water waves, with conformal symmetry being only one of the potential symmetries which could be leveraged, in order to realize extremely flat bands for the localization of waves.

\section*{Acknowledgments}
The authors thank Prof. Ortwin Hess, Dr. Freddie Page and Dr. Fan Yang for helpful discussions. This work was supported by Natural Science Foundation of China (NSFC) (51977165, 51507123), the Fundamental Research Funds for the Central
Universities (xjj2018189), China Scholarship Council (201806280476), Engineering and Physical Sciences Research Council (EPSRC) (EP/L015579/1), and Gordon and Betty Moore Foundation.
\bibliography{plasmonic_loc-Lu}

\providecommand{\noopsort}[1]{}\providecommand{\singleletter}[1]{#1}%
\begin{thebibliography}{44}%
\makeatletter
\providecommand \@ifxundefined [1]{%
 \@ifx{#1\undefined}
}%
\providecommand \@ifnum [1]{%
 \ifnum #1\expandafter \@firstoftwo
 \else \expandafter \@secondoftwo
 \fi
}%
\providecommand \@ifx [1]{%
 \ifx #1\expandafter \@firstoftwo
 \else \expandafter \@secondoftwo
 \fi
}%
\providecommand \natexlab [1]{#1}%
\providecommand \enquote  [1]{``#1''}%
\providecommand \bibnamefont  [1]{#1}%
\providecommand \bibfnamefont [1]{#1}%
\providecommand \citenamefont [1]{#1}%
\providecommand \href@noop [0]{\@secondoftwo}%
\providecommand \href [0]{\begingroup \@sanitize@url \@href}%
\providecommand \@href[1]{\@@startlink{#1}\@@href}%
\providecommand \@@href[1]{\endgroup#1\@@endlink}%
\providecommand \@sanitize@url [0]{\catcode `\\12\catcode `\$12\catcode
  `\&12\catcode `\#12\catcode `\^12\catcode `\_12\catcode `\%12\relax}%
\providecommand \@@startlink[1]{}%
\providecommand \@@endlink[0]{}%
\providecommand \url  [0]{\begingroup\@sanitize@url \@url }%
\providecommand \@url [1]{\endgroup\@href {#1}{\urlprefix }}%
\providecommand \urlprefix  [0]{URL }%
\providecommand \Eprint [0]{\href }%
\providecommand \doibase [0]{http://dx.doi.org/}%
\providecommand \selectlanguage [0]{\@gobble}%
\providecommand \bibinfo  [0]{\@secondoftwo}%
\providecommand \bibfield  [0]{\@secondoftwo}%
\providecommand \translation [1]{[#1]}%
\providecommand \BibitemOpen [0]{}%
\providecommand \bibitemStop [0]{}%
\providecommand \bibitemNoStop [0]{.\EOS\space}%
\providecommand \EOS [0]{\spacefactor3000\relax}%
\providecommand \BibitemShut  [1]{\csname bibitem#1\endcsname}%
\let\auto@bib@innerbib\@empty
\bibitem [{\citenamefont {Dresselhaus}\ \emph {et~al.}(2008)\citenamefont
  {Dresselhaus}, \citenamefont {Dresselhaus},\ and\ \citenamefont
  {Jorio}}]{dresselhaus2008applications}%
  \BibitemOpen
  \bibfield  {author} {\bibinfo {author} {\bibfnamefont {M.~S.}\ \bibnamefont
  {Dresselhaus}}, \bibinfo {author} {\bibfnamefont {G.}~\bibnamefont
  {Dresselhaus}}, \ and\ \bibinfo {author} {\bibfnamefont {A.}~\bibnamefont
  {Jorio}},\ }\href@noop {} {\enquote {\bibinfo {title} {Applications of group
  theory to the physics of solids},}\ } (\bibinfo {year} {2008})\BibitemShut
  {NoStop}%
\bibitem [{\citenamefont {Sakoda}(2004)}]{sakoda2004optical}%
  \BibitemOpen
  \bibfield  {author} {\bibinfo {author} {\bibfnamefont {K.}~\bibnamefont
  {Sakoda}},\ }\href@noop {} {\emph {\bibinfo {title} {Optical properties of
  photonic crystals}}},\ Vol.~\bibinfo {volume} {80}\ (\bibinfo  {publisher}
  {Springer Science \& Business Media},\ \bibinfo {year} {2004})\BibitemShut
  {NoStop}%
\bibitem [{\citenamefont {Ward}\ and\ \citenamefont
  {Pendry}(1996)}]{ward1996refraction}%
  \BibitemOpen
  \bibfield  {author} {\bibinfo {author} {\bibfnamefont {A.}~\bibnamefont
  {Ward}}\ and\ \bibinfo {author} {\bibfnamefont {J.~B.}\ \bibnamefont
  {Pendry}},\ }\href@noop {} {\bibfield  {journal} {\bibinfo  {journal}
  {Journal of Modern Optics}\ }\textbf {\bibinfo {volume} {43}},\ \bibinfo
  {pages} {773} (\bibinfo {year} {1996})}\BibitemShut {NoStop}%
\bibitem [{\citenamefont {Pendry}\ \emph {et~al.}(2006)\citenamefont {Pendry},
  \citenamefont {Schurig},\ and\ \citenamefont
  {Smith}}]{pendry2006controlling}%
  \BibitemOpen
  \bibfield  {author} {\bibinfo {author} {\bibfnamefont {J.~B.}\ \bibnamefont
  {Pendry}}, \bibinfo {author} {\bibfnamefont {D.}~\bibnamefont {Schurig}}, \
  and\ \bibinfo {author} {\bibfnamefont {D.~R.}\ \bibnamefont {Smith}},\
  }\href@noop {} {\bibfield  {journal} {\bibinfo  {journal} {Science}\ }\textbf
  {\bibinfo {volume} {312}},\ \bibinfo {pages} {1780} (\bibinfo {year}
  {2006})}\BibitemShut {NoStop}%
\bibitem [{\citenamefont {Leonhardt}(2006)}]{leonhardt2006optical}%
  \BibitemOpen
  \bibfield  {author} {\bibinfo {author} {\bibfnamefont {U.}~\bibnamefont
  {Leonhardt}},\ }\href@noop {} {\bibfield  {journal} {\bibinfo  {journal}
  {Science}\ }\textbf {\bibinfo {volume} {312}},\ \bibinfo {pages} {1777}
  (\bibinfo {year} {2006})}\BibitemShut {NoStop}%
\bibitem [{\citenamefont {Huidobro}\ \emph {et~al.}(2010)\citenamefont
  {Huidobro}, \citenamefont {Nesterov}, \citenamefont {Mart{\'\i}n-Moreno},\
  and\ \citenamefont {Garc{\'\i}a-Vidal}}]{huidobro2010transformation}%
  \BibitemOpen
  \bibfield  {author} {\bibinfo {author} {\bibfnamefont {P.~A.}\ \bibnamefont
  {Huidobro}}, \bibinfo {author} {\bibfnamefont {M.~L.}\ \bibnamefont
  {Nesterov}}, \bibinfo {author} {\bibfnamefont {L.}~\bibnamefont
  {Mart{\'\i}n-Moreno}}, \ and\ \bibinfo {author} {\bibfnamefont {F.~J.}\
  \bibnamefont {Garc{\'\i}a-Vidal}},\ }\href@noop {} {\bibfield  {journal}
  {\bibinfo  {journal} {Nano Letters}\ }\textbf {\bibinfo {volume} {10}},\
  \bibinfo {pages} {1985} (\bibinfo {year} {2010})}\BibitemShut {NoStop}%
\bibitem [{\citenamefont {Pendry}\ \emph {et~al.}(2012)\citenamefont {Pendry},
  \citenamefont {Aubry}, \citenamefont {Smith},\ and\ \citenamefont
  {Maier}}]{pendry2012transformation}%
  \BibitemOpen
  \bibfield  {author} {\bibinfo {author} {\bibfnamefont {J.}~\bibnamefont
  {Pendry}}, \bibinfo {author} {\bibfnamefont {A.}~\bibnamefont {Aubry}},
  \bibinfo {author} {\bibfnamefont {D.}~\bibnamefont {Smith}}, \ and\ \bibinfo
  {author} {\bibfnamefont {S.}~\bibnamefont {Maier}},\ }\href@noop {}
  {\bibfield  {journal} {\bibinfo  {journal} {Science}\ }\textbf {\bibinfo
  {volume} {337}},\ \bibinfo {pages} {549} (\bibinfo {year}
  {2012})}\BibitemShut {NoStop}%
\bibitem [{\citenamefont {Kraft}\ \emph {et~al.}(2015)\citenamefont {Kraft},
  \citenamefont {Luo}, \citenamefont {Maier},\ and\ \citenamefont
  {Pendry}}]{kraft2015designing}%
  \BibitemOpen
  \bibfield  {author} {\bibinfo {author} {\bibfnamefont {M.}~\bibnamefont
  {Kraft}}, \bibinfo {author} {\bibfnamefont {Y.}~\bibnamefont {Luo}}, \bibinfo
  {author} {\bibfnamefont {S.}~\bibnamefont {Maier}}, \ and\ \bibinfo {author}
  {\bibfnamefont {J.}~\bibnamefont {Pendry}},\ }\href@noop {} {\bibfield
  {journal} {\bibinfo  {journal} {Physical Review X}\ }\textbf {\bibinfo
  {volume} {5}},\ \bibinfo {pages} {031029} (\bibinfo {year}
  {2015})}\BibitemShut {NoStop}%
\bibitem [{\citenamefont {Kraft}\ \emph {et~al.}(2014)\citenamefont {Kraft},
  \citenamefont {Pendry}, \citenamefont {Maier},\ and\ \citenamefont
  {Luo}}]{kraft2014transformation}%
  \BibitemOpen
  \bibfield  {author} {\bibinfo {author} {\bibfnamefont {M.}~\bibnamefont
  {Kraft}}, \bibinfo {author} {\bibfnamefont {J.}~\bibnamefont {Pendry}},
  \bibinfo {author} {\bibfnamefont {S.}~\bibnamefont {Maier}}, \ and\ \bibinfo
  {author} {\bibfnamefont {Y.}~\bibnamefont {Luo}},\ }\href@noop {} {\bibfield
  {journal} {\bibinfo  {journal} {Physical Review B}\ }\textbf {\bibinfo
  {volume} {89}},\ \bibinfo {pages} {245125} (\bibinfo {year}
  {2014})}\BibitemShut {NoStop}%
\bibitem [{\citenamefont {Hau}\ \emph {et~al.}(1999)\citenamefont {Hau},
  \citenamefont {Harris}, \citenamefont {Dutton},\ and\ \citenamefont
  {Behroozi}}]{hau1999light}%
  \BibitemOpen
  \bibfield  {author} {\bibinfo {author} {\bibfnamefont {L.~V.}\ \bibnamefont
  {Hau}}, \bibinfo {author} {\bibfnamefont {S.~E.}\ \bibnamefont {Harris}},
  \bibinfo {author} {\bibfnamefont {Z.}~\bibnamefont {Dutton}}, \ and\ \bibinfo
  {author} {\bibfnamefont {C.~H.}\ \bibnamefont {Behroozi}},\ }\href@noop {}
  {\bibfield  {journal} {\bibinfo  {journal} {Nature}\ }\textbf {\bibinfo
  {volume} {397}},\ \bibinfo {pages} {594} (\bibinfo {year}
  {1999})}\BibitemShut {NoStop}%
\bibitem [{\citenamefont {Boyd}\ and\ \citenamefont
  {Gauthier}(2009)}]{boyd2009controlling}%
  \BibitemOpen
  \bibfield  {author} {\bibinfo {author} {\bibfnamefont {R.~W.}\ \bibnamefont
  {Boyd}}\ and\ \bibinfo {author} {\bibfnamefont {D.~J.}\ \bibnamefont
  {Gauthier}},\ }\href@noop {} {\bibfield  {journal} {\bibinfo  {journal}
  {Science}\ }\textbf {\bibinfo {volume} {326}},\ \bibinfo {pages} {1074}
  (\bibinfo {year} {2009})}\BibitemShut {NoStop}%
\bibitem [{\citenamefont {Settle}\ \emph {et~al.}(2007)\citenamefont {Settle},
  \citenamefont {Engelen}, \citenamefont {Salib}, \citenamefont {Michaeli},
  \citenamefont {Kuipers},\ and\ \citenamefont {Krauss}}]{settle2007flatband}%
  \BibitemOpen
  \bibfield  {author} {\bibinfo {author} {\bibfnamefont {M.}~\bibnamefont
  {Settle}}, \bibinfo {author} {\bibfnamefont {R.}~\bibnamefont {Engelen}},
  \bibinfo {author} {\bibfnamefont {M.}~\bibnamefont {Salib}}, \bibinfo
  {author} {\bibfnamefont {A.}~\bibnamefont {Michaeli}}, \bibinfo {author}
  {\bibfnamefont {L.}~\bibnamefont {Kuipers}}, \ and\ \bibinfo {author}
  {\bibfnamefont {T.}~\bibnamefont {Krauss}},\ }\href@noop {} {\bibfield
  {journal} {\bibinfo  {journal} {Optics Express}\ }\textbf {\bibinfo {volume}
  {15}},\ \bibinfo {pages} {219} (\bibinfo {year} {2007})}\BibitemShut
  {NoStop}%
\bibitem [{\citenamefont {Baba}(2008)}]{baba2008slow}%
  \BibitemOpen
  \bibfield  {author} {\bibinfo {author} {\bibfnamefont {T.}~\bibnamefont
  {Baba}},\ }\href@noop {} {\bibfield  {journal} {\bibinfo  {journal} {Nature
  Photonics}\ }\textbf {\bibinfo {volume} {2}},\ \bibinfo {pages} {465}
  (\bibinfo {year} {2008})}\BibitemShut {NoStop}%
\bibitem [{\citenamefont {Yu}\ and\ \citenamefont
  {Capasso}(2014)}]{yu2014flat}%
  \BibitemOpen
  \bibfield  {author} {\bibinfo {author} {\bibfnamefont {N.}~\bibnamefont
  {Yu}}\ and\ \bibinfo {author} {\bibfnamefont {F.}~\bibnamefont {Capasso}},\
  }\href@noop {} {\bibfield  {journal} {\bibinfo  {journal} {Nature Materials}\
  }\textbf {\bibinfo {volume} {13}},\ \bibinfo {pages} {139} (\bibinfo {year}
  {2014})}\BibitemShut {NoStop}%
\bibitem [{\citenamefont {Kildishev}\ \emph {et~al.}(2013)\citenamefont
  {Kildishev}, \citenamefont {Boltasseva},\ and\ \citenamefont
  {Shalaev}}]{kildishev2013planar}%
  \BibitemOpen
  \bibfield  {author} {\bibinfo {author} {\bibfnamefont {A.~V.}\ \bibnamefont
  {Kildishev}}, \bibinfo {author} {\bibfnamefont {A.}~\bibnamefont
  {Boltasseva}}, \ and\ \bibinfo {author} {\bibfnamefont {V.~M.}\ \bibnamefont
  {Shalaev}},\ }\href@noop {} {\bibfield  {journal} {\bibinfo  {journal}
  {Science}\ }\textbf {\bibinfo {volume} {339}},\ \bibinfo {pages} {1232009}
  (\bibinfo {year} {2013})}\BibitemShut {NoStop}%
\bibitem [{\citenamefont {Shaltout}\ \emph {et~al.}(2019)\citenamefont
  {Shaltout}, \citenamefont {Shalaev},\ and\ \citenamefont
  {Brongersma}}]{shaltout2019spatiotemporal}%
  \BibitemOpen
  \bibfield  {author} {\bibinfo {author} {\bibfnamefont {A.~M.}\ \bibnamefont
  {Shaltout}}, \bibinfo {author} {\bibfnamefont {V.~M.}\ \bibnamefont
  {Shalaev}}, \ and\ \bibinfo {author} {\bibfnamefont {M.~L.}\ \bibnamefont
  {Brongersma}},\ }\href@noop {} {\bibfield  {journal} {\bibinfo  {journal}
  {Science}\ }\textbf {\bibinfo {volume} {364}},\ \bibinfo {pages} {eaat3100}
  (\bibinfo {year} {2019})}\BibitemShut {NoStop}%
\bibitem [{\citenamefont {Tsakmakidis}\ \emph {et~al.}(2007)\citenamefont
  {Tsakmakidis}, \citenamefont {Boardman},\ and\ \citenamefont
  {Hess}}]{tsakmakidis2007trapped}%
  \BibitemOpen
  \bibfield  {author} {\bibinfo {author} {\bibfnamefont {K.~L.}\ \bibnamefont
  {Tsakmakidis}}, \bibinfo {author} {\bibfnamefont {A.~D.}\ \bibnamefont
  {Boardman}}, \ and\ \bibinfo {author} {\bibfnamefont {O.}~\bibnamefont
  {Hess}},\ }\href@noop {} {\bibfield  {journal} {\bibinfo  {journal} {Nature}\
  }\textbf {\bibinfo {volume} {450}},\ \bibinfo {pages} {397} (\bibinfo {year}
  {2007})}\BibitemShut {NoStop}%
\bibitem [{\citenamefont {Tsakmakidis}\ \emph {et~al.}(2014)\citenamefont
  {Tsakmakidis}, \citenamefont {Pickering}, \citenamefont {Hamm}, \citenamefont
  {Page},\ and\ \citenamefont {Hess}}]{tsakmakidis2014completely}%
  \BibitemOpen
  \bibfield  {author} {\bibinfo {author} {\bibfnamefont {K.~L.}\ \bibnamefont
  {Tsakmakidis}}, \bibinfo {author} {\bibfnamefont {T.~W.}\ \bibnamefont
  {Pickering}}, \bibinfo {author} {\bibfnamefont {J.~M.}\ \bibnamefont {Hamm}},
  \bibinfo {author} {\bibfnamefont {A.~F.}\ \bibnamefont {Page}}, \ and\
  \bibinfo {author} {\bibfnamefont {O.}~\bibnamefont {Hess}},\ }\href@noop {}
  {\bibfield  {journal} {\bibinfo  {journal} {Phys. Rev. Lett.}\ }\textbf
  {\bibinfo {volume} {112}},\ \bibinfo {pages} {167401} (\bibinfo {year}
  {2014})}\BibitemShut {NoStop}%
\bibitem [{\citenamefont {Tsakmakidis}\ \emph {et~al.}(2017)\citenamefont
  {Tsakmakidis}, \citenamefont {Hess}, \citenamefont {Boyd},\ and\
  \citenamefont {Zhang}}]{tsakmakidis2017ultraslow}%
  \BibitemOpen
  \bibfield  {author} {\bibinfo {author} {\bibfnamefont {K.~L.}\ \bibnamefont
  {Tsakmakidis}}, \bibinfo {author} {\bibfnamefont {O.}~\bibnamefont {Hess}},
  \bibinfo {author} {\bibfnamefont {R.~W.}\ \bibnamefont {Boyd}}, \ and\
  \bibinfo {author} {\bibfnamefont {X.}~\bibnamefont {Zhang}},\ }\href@noop {}
  {\bibfield  {journal} {\bibinfo  {journal} {Science}\ }\textbf {\bibinfo
  {volume} {358}},\ \bibinfo {pages} {eaan5196} (\bibinfo {year}
  {2017})}\BibitemShut {NoStop}%
\bibitem [{\citenamefont {Gan}\ \emph {et~al.}(2008)\citenamefont {Gan},
  \citenamefont {Fu}, \citenamefont {Ding},\ and\ \citenamefont
  {Bartoli}}]{gan2008ultrawide}%
  \BibitemOpen
  \bibfield  {author} {\bibinfo {author} {\bibfnamefont {Q.}~\bibnamefont
  {Gan}}, \bibinfo {author} {\bibfnamefont {Z.}~\bibnamefont {Fu}}, \bibinfo
  {author} {\bibfnamefont {Y.~J.}\ \bibnamefont {Ding}}, \ and\ \bibinfo
  {author} {\bibfnamefont {F.~J.}\ \bibnamefont {Bartoli}},\ }\href@noop {}
  {\bibfield  {journal} {\bibinfo  {journal} {Physical Review Letters}\
  }\textbf {\bibinfo {volume} {100}},\ \bibinfo {pages} {256803} (\bibinfo
  {year} {2008})}\BibitemShut {NoStop}%
\bibitem [{\citenamefont {Hao}\ \emph {et~al.}(2019)\citenamefont {Hao},
  \citenamefont {Ye}, \citenamefont {Jiao},\ and\ \citenamefont
  {Li}}]{hao2019increasing}%
  \BibitemOpen
  \bibfield  {author} {\bibinfo {author} {\bibfnamefont {R.}~\bibnamefont
  {Hao}}, \bibinfo {author} {\bibfnamefont {G.}~\bibnamefont {Ye}}, \bibinfo
  {author} {\bibfnamefont {J.}~\bibnamefont {Jiao}}, \ and\ \bibinfo {author}
  {\bibfnamefont {E.}~\bibnamefont {Li}},\ }\href@noop {} {\bibfield  {journal}
  {\bibinfo  {journal} {Photonics Research}\ }\textbf {\bibinfo {volume} {7}},\
  \bibinfo {pages} {240} (\bibinfo {year} {2019})}\BibitemShut {NoStop}%
\bibitem [{\citenamefont {Kubo}\ \emph {et~al.}(2007)\citenamefont {Kubo},
  \citenamefont {Mori},\ and\ \citenamefont {Baba}}]{kubo2007low}%
  \BibitemOpen
  \bibfield  {author} {\bibinfo {author} {\bibfnamefont {S.}~\bibnamefont
  {Kubo}}, \bibinfo {author} {\bibfnamefont {D.}~\bibnamefont {Mori}}, \ and\
  \bibinfo {author} {\bibfnamefont {T.}~\bibnamefont {Baba}},\ }\href@noop {}
  {\bibfield  {journal} {\bibinfo  {journal} {Optics Letters}\ }\textbf
  {\bibinfo {volume} {32}},\ \bibinfo {pages} {2981} (\bibinfo {year}
  {2007})}\BibitemShut {NoStop}%
\bibitem [{\citenamefont {Minkov}\ and\ \citenamefont
  {Savona}(2015)}]{minkov2015wide}%
  \BibitemOpen
  \bibfield  {author} {\bibinfo {author} {\bibfnamefont {M.}~\bibnamefont
  {Minkov}}\ and\ \bibinfo {author} {\bibfnamefont {V.}~\bibnamefont
  {Savona}},\ }\href@noop {} {\bibfield  {journal} {\bibinfo  {journal}
  {Optica}\ }\textbf {\bibinfo {volume} {2}},\ \bibinfo {pages} {631} (\bibinfo
  {year} {2015})}\BibitemShut {NoStop}%
\bibitem [{\citenamefont {Hao}\ \emph {et~al.}(2010)\citenamefont {Hao},
  \citenamefont {Cassan}, \citenamefont {Kurt}, \citenamefont {Le~Roux},
  \citenamefont {Marris-Morini}, \citenamefont {Vivien}, \citenamefont {Wu},
  \citenamefont {Zhou},\ and\ \citenamefont {Zhang}}]{hao2010novel}%
  \BibitemOpen
  \bibfield  {author} {\bibinfo {author} {\bibfnamefont {R.}~\bibnamefont
  {Hao}}, \bibinfo {author} {\bibfnamefont {E.}~\bibnamefont {Cassan}},
  \bibinfo {author} {\bibfnamefont {H.}~\bibnamefont {Kurt}}, \bibinfo {author}
  {\bibfnamefont {X.}~\bibnamefont {Le~Roux}}, \bibinfo {author} {\bibfnamefont
  {D.}~\bibnamefont {Marris-Morini}}, \bibinfo {author} {\bibfnamefont
  {L.}~\bibnamefont {Vivien}}, \bibinfo {author} {\bibfnamefont
  {H.}~\bibnamefont {Wu}}, \bibinfo {author} {\bibfnamefont {Z.}~\bibnamefont
  {Zhou}}, \ and\ \bibinfo {author} {\bibfnamefont {X.}~\bibnamefont {Zhang}},\
  }\href@noop {} {\bibfield  {journal} {\bibinfo  {journal} {Optics Express}\
  }\textbf {\bibinfo {volume} {18}},\ \bibinfo {pages} {5942} (\bibinfo {year}
  {2010})}\BibitemShut {NoStop}%
\bibitem [{\citenamefont {Woessner}\ \emph {et~al.}(2015)\citenamefont
  {Woessner}, \citenamefont {Lundeberg}, \citenamefont {Gao}, \citenamefont
  {Principi}, \citenamefont {Alonso-Gonz{\'a}lez}, \citenamefont {Carrega},
  \citenamefont {Watanabe}, \citenamefont {Taniguchi}, \citenamefont {Vignale},
  \citenamefont {Polini} \emph {et~al.}}]{woessner2015highly}%
  \BibitemOpen
  \bibfield  {author} {\bibinfo {author} {\bibfnamefont {A.}~\bibnamefont
  {Woessner}}, \bibinfo {author} {\bibfnamefont {M.~B.}\ \bibnamefont
  {Lundeberg}}, \bibinfo {author} {\bibfnamefont {Y.}~\bibnamefont {Gao}},
  \bibinfo {author} {\bibfnamefont {A.}~\bibnamefont {Principi}}, \bibinfo
  {author} {\bibfnamefont {P.}~\bibnamefont {Alonso-Gonz{\'a}lez}}, \bibinfo
  {author} {\bibfnamefont {M.}~\bibnamefont {Carrega}}, \bibinfo {author}
  {\bibfnamefont {K.}~\bibnamefont {Watanabe}}, \bibinfo {author}
  {\bibfnamefont {T.}~\bibnamefont {Taniguchi}}, \bibinfo {author}
  {\bibfnamefont {G.}~\bibnamefont {Vignale}}, \bibinfo {author} {\bibfnamefont
  {M.}~\bibnamefont {Polini}},  \emph {et~al.},\ }\href@noop {} {\bibfield
  {journal} {\bibinfo  {journal} {Nature Materials}\ }\textbf {\bibinfo
  {volume} {14}},\ \bibinfo {pages} {421} (\bibinfo {year} {2015})}\BibitemShut
  {NoStop}%
\bibitem [{\citenamefont {Chubchev}\ \emph {et~al.}(2018)\citenamefont
  {Chubchev}, \citenamefont {Nechepurenko}, \citenamefont {Dorofeenko},
  \citenamefont {Vinogradov},\ and\ \citenamefont
  {Lisyansky}}]{chubchev2018highly}%
  \BibitemOpen
  \bibfield  {author} {\bibinfo {author} {\bibfnamefont {E.}~\bibnamefont
  {Chubchev}}, \bibinfo {author} {\bibfnamefont {I.}~\bibnamefont
  {Nechepurenko}}, \bibinfo {author} {\bibfnamefont {A.}~\bibnamefont
  {Dorofeenko}}, \bibinfo {author} {\bibfnamefont {A.}~\bibnamefont
  {Vinogradov}}, \ and\ \bibinfo {author} {\bibfnamefont {A.}~\bibnamefont
  {Lisyansky}},\ }\href@noop {} {\bibfield  {journal} {\bibinfo  {journal}
  {Optics Express}\ }\textbf {\bibinfo {volume} {26}},\ \bibinfo {pages} {9050}
  (\bibinfo {year} {2018})}\BibitemShut {NoStop}%
\bibitem [{\citenamefont {Kauranen}\ and\ \citenamefont
  {Zayats}(2012)}]{kauranen2012nonlinear}%
  \BibitemOpen
  \bibfield  {author} {\bibinfo {author} {\bibfnamefont {M.}~\bibnamefont
  {Kauranen}}\ and\ \bibinfo {author} {\bibfnamefont {A.~V.}\ \bibnamefont
  {Zayats}},\ }\href@noop {} {\bibfield  {journal} {\bibinfo  {journal} {Nature
  Photonics}\ }\textbf {\bibinfo {volume} {6}},\ \bibinfo {pages} {737}
  (\bibinfo {year} {2012})}\BibitemShut {NoStop}%
\bibitem [{\citenamefont {Luo}\ \emph {et~al.}(2013)\citenamefont {Luo},
  \citenamefont {Zhao}, \citenamefont {Fernandez-Dominguez}, \citenamefont
  {Maier},\ and\ \citenamefont {Pendry}}]{luo2013harvesting}%
  \BibitemOpen
  \bibfield  {author} {\bibinfo {author} {\bibfnamefont {Y.}~\bibnamefont
  {Luo}}, \bibinfo {author} {\bibfnamefont {R.}~\bibnamefont {Zhao}}, \bibinfo
  {author} {\bibfnamefont {A.~I.}\ \bibnamefont {Fernandez-Dominguez}},
  \bibinfo {author} {\bibfnamefont {S.~A.}\ \bibnamefont {Maier}}, \ and\
  \bibinfo {author} {\bibfnamefont {J.~B.}\ \bibnamefont {Pendry}},\
  }\href@noop {} {\bibfield  {journal} {\bibinfo  {journal} {Science China
  Information Sciences}\ }\textbf {\bibinfo {volume} {56}},\ \bibinfo {pages}
  {1} (\bibinfo {year} {2013})}\BibitemShut {NoStop}%
\bibitem [{\citenamefont {Chon}\ \emph {et~al.}(2007)\citenamefont {Chon},
  \citenamefont {Bullen}, \citenamefont {Zijlstra},\ and\ \citenamefont
  {Gu}}]{chon2007spectral}%
  \BibitemOpen
  \bibfield  {author} {\bibinfo {author} {\bibfnamefont {J.~W.}\ \bibnamefont
  {Chon}}, \bibinfo {author} {\bibfnamefont {C.}~\bibnamefont {Bullen}},
  \bibinfo {author} {\bibfnamefont {P.}~\bibnamefont {Zijlstra}}, \ and\
  \bibinfo {author} {\bibfnamefont {M.}~\bibnamefont {Gu}},\ }\href@noop {}
  {\bibfield  {journal} {\bibinfo  {journal} {Advanced Functional Materials}\
  }\textbf {\bibinfo {volume} {17}},\ \bibinfo {pages} {875} (\bibinfo {year}
  {2007})}\BibitemShut {NoStop}%
\bibitem [{\citenamefont {Kongsuwan}\ \emph {et~al.}(2019)\citenamefont
  {Kongsuwan}, \citenamefont {Xiong}, \citenamefont {Bai}, \citenamefont {You},
  \citenamefont {Png}, \citenamefont {Wu},\ and\ \citenamefont
  {Hess}}]{kongsuwan2019quantum}%
  \BibitemOpen
  \bibfield  {author} {\bibinfo {author} {\bibfnamefont {N.}~\bibnamefont
  {Kongsuwan}}, \bibinfo {author} {\bibfnamefont {X.}~\bibnamefont {Xiong}},
  \bibinfo {author} {\bibfnamefont {P.}~\bibnamefont {Bai}}, \bibinfo {author}
  {\bibfnamefont {J.-B.}\ \bibnamefont {You}}, \bibinfo {author} {\bibfnamefont
  {C.~E.}\ \bibnamefont {Png}}, \bibinfo {author} {\bibfnamefont
  {L.}~\bibnamefont {Wu}}, \ and\ \bibinfo {author} {\bibfnamefont
  {O.}~\bibnamefont {Hess}},\ }\href@noop {} {\bibfield  {journal} {\bibinfo
  {journal} {Nano Letters}\ }\textbf {\bibinfo {volume} {19}},\ \bibinfo
  {pages} {5853} (\bibinfo {year} {2019})}\BibitemShut {NoStop}%
\bibitem [{\citenamefont {Mansuripur}\ \emph {et~al.}(2009)\citenamefont
  {Mansuripur}, \citenamefont {Zakharian}, \citenamefont {Lesuffleur},
  \citenamefont {Oh}, \citenamefont {Jones}, \citenamefont {Lindquist},
  \citenamefont {Im}, \citenamefont {Kobyakov},\ and\ \citenamefont
  {Moloney}}]{mansuripur2009plasmonic}%
  \BibitemOpen
  \bibfield  {author} {\bibinfo {author} {\bibfnamefont {M.}~\bibnamefont
  {Mansuripur}}, \bibinfo {author} {\bibfnamefont {A.~R.}\ \bibnamefont
  {Zakharian}}, \bibinfo {author} {\bibfnamefont {A.}~\bibnamefont
  {Lesuffleur}}, \bibinfo {author} {\bibfnamefont {S.-H.}\ \bibnamefont {Oh}},
  \bibinfo {author} {\bibfnamefont {R.~J.}\ \bibnamefont {Jones}}, \bibinfo
  {author} {\bibfnamefont {N.}~\bibnamefont {Lindquist}}, \bibinfo {author}
  {\bibfnamefont {H.}~\bibnamefont {Im}}, \bibinfo {author} {\bibfnamefont
  {A.}~\bibnamefont {Kobyakov}}, \ and\ \bibinfo {author} {\bibfnamefont
  {J.~V.}\ \bibnamefont {Moloney}},\ }\href@noop {} {\bibfield  {journal}
  {\bibinfo  {journal} {Optics Express}\ }\textbf {\bibinfo {volume} {17}},\
  \bibinfo {pages} {14001} (\bibinfo {year} {2009})}\BibitemShut {NoStop}%
\bibitem [{\citenamefont {Luo}\ \emph {et~al.}(2012)\citenamefont {Luo},
  \citenamefont {Lei}, \citenamefont {Maier},\ and\ \citenamefont
  {Pendry}}]{luo2012broadband}%
  \BibitemOpen
  \bibfield  {author} {\bibinfo {author} {\bibfnamefont {Y.}~\bibnamefont
  {Luo}}, \bibinfo {author} {\bibfnamefont {D.~Y.}\ \bibnamefont {Lei}},
  \bibinfo {author} {\bibfnamefont {S.~A.}\ \bibnamefont {Maier}}, \ and\
  \bibinfo {author} {\bibfnamefont {J.}~\bibnamefont {Pendry}},\ }\href@noop {}
  {\bibfield  {journal} {\bibinfo  {journal} {Physical Review Letters}\
  }\textbf {\bibinfo {volume} {108}},\ \bibinfo {pages} {023901} (\bibinfo
  {year} {2012})}\BibitemShut {NoStop}%
\bibitem [{\citenamefont {Schinzinger}\ and\ \citenamefont
  {Laura}(2012)}]{schinzinger2012conformal}%
  \BibitemOpen
  \bibfield  {author} {\bibinfo {author} {\bibfnamefont {R.}~\bibnamefont
  {Schinzinger}}\ and\ \bibinfo {author} {\bibfnamefont {P.~A.}\ \bibnamefont
  {Laura}},\ }\href@noop {} {\emph {\bibinfo {title} {Conformal mapping:
  methods and applications}}}\ (\bibinfo  {publisher} {Courier Corporation},\
  \bibinfo {year} {2012})\BibitemShut {NoStop}%
\bibitem [{\citenamefont {Pendry}\ \emph {et~al.}(2019)\citenamefont {Pendry},
  \citenamefont {Huidobro},\ and\ \citenamefont {Ding}}]{pendry2019computing}%
  \BibitemOpen
  \bibfield  {author} {\bibinfo {author} {\bibfnamefont {J.}~\bibnamefont
  {Pendry}}, \bibinfo {author} {\bibfnamefont {P.~A.}\ \bibnamefont
  {Huidobro}}, \ and\ \bibinfo {author} {\bibfnamefont {K.}~\bibnamefont
  {Ding}},\ }\href@noop {} {\bibfield  {journal} {\bibinfo  {journal} {Physical
  Review B}\ }\textbf {\bibinfo {volume} {99}},\ \bibinfo {pages} {085408}
  (\bibinfo {year} {2019})}\BibitemShut {NoStop}%
\bibitem [{\citenamefont {Cui}\ \emph {et~al.}(2019)\citenamefont {Cui},
  \citenamefont {Ding}, \citenamefont {Dong},\ and\ \citenamefont
  {Chan}}]{PhysRevB.100.115412}%
  \BibitemOpen
  \bibfield  {author} {\bibinfo {author} {\bibfnamefont {X.}~\bibnamefont
  {Cui}}, \bibinfo {author} {\bibfnamefont {K.}~\bibnamefont {Ding}}, \bibinfo
  {author} {\bibfnamefont {J.-W.}\ \bibnamefont {Dong}}, \ and\ \bibinfo
  {author} {\bibfnamefont {C.~T.}\ \bibnamefont {Chan}},\ }\href {\doibase
  10.1103/PhysRevB.100.115412} {\bibfield  {journal} {\bibinfo  {journal}
  {Phys. Rev. B}\ }\textbf {\bibinfo {volume} {100}},\ \bibinfo {pages}
  {115412} (\bibinfo {year} {2019})}\BibitemShut {NoStop}%
\bibitem [{\citenamefont {Maier}(2007)}]{maier2007plasmonics}%
  \BibitemOpen
  \bibfield  {author} {\bibinfo {author} {\bibfnamefont {S.~A.}\ \bibnamefont
  {Maier}},\ }\href@noop {} {\emph {\bibinfo {title} {Plasmonics: fundamentals
  and applications}}}\ (\bibinfo  {publisher} {Springer Science \& Business
  Media},\ \bibinfo {year} {2007})\BibitemShut {NoStop}%
\bibitem [{\citenamefont {Jackson}(1999)}]{jackson1999classical}%
  \BibitemOpen
  \bibfield  {author} {\bibinfo {author} {\bibfnamefont {J.~D.}\ \bibnamefont
  {Jackson}},\ }\href@noop {} {\enquote {\bibinfo {title} {Classical
  electrodynamics},}\ } (\bibinfo {year} {1999})\BibitemShut {NoStop}%
\bibitem [{\citenamefont {Pendry}\ \emph {et~al.}(2017)\citenamefont {Pendry},
  \citenamefont {Huidobro}, \citenamefont {Luo},\ and\ \citenamefont
  {Galiffi}}]{pendry2017compacted}%
  \BibitemOpen
  \bibfield  {author} {\bibinfo {author} {\bibfnamefont {J.}~\bibnamefont
  {Pendry}}, \bibinfo {author} {\bibfnamefont {P.~A.}\ \bibnamefont
  {Huidobro}}, \bibinfo {author} {\bibfnamefont {Y.}~\bibnamefont {Luo}}, \
  and\ \bibinfo {author} {\bibfnamefont {E.}~\bibnamefont {Galiffi}},\
  }\href@noop {} {\bibfield  {journal} {\bibinfo  {journal} {Science}\ }\textbf
  {\bibinfo {volume} {358}},\ \bibinfo {pages} {915} (\bibinfo {year}
  {2017})}\BibitemShut {NoStop}%
\bibitem [{\citenamefont {Galiffi}\ \emph {et~al.}(2019)\citenamefont
  {Galiffi}, \citenamefont {Pendry},\ and\ \citenamefont
  {Huidobro}}]{galiffi2019singular}%
  \BibitemOpen
  \bibfield  {author} {\bibinfo {author} {\bibfnamefont {E.}~\bibnamefont
  {Galiffi}}, \bibinfo {author} {\bibfnamefont {J.}~\bibnamefont {Pendry}}, \
  and\ \bibinfo {author} {\bibfnamefont {P.~A.}\ \bibnamefont {Huidobro}},\
  }\href@noop {} {\bibfield  {journal} {\bibinfo  {journal} {EPJ Applied
  Metamaterials}\ }\textbf {\bibinfo {volume} {6}},\ \bibinfo {pages} {10}
  (\bibinfo {year} {2019})}\BibitemShut {NoStop}%
\bibitem [{\citenamefont {Yang}\ \emph {et~al.}(2018)\citenamefont {Yang},
  \citenamefont {Huidobro},\ and\ \citenamefont
  {Pendry}}]{yang2018transformation}%
  \BibitemOpen
  \bibfield  {author} {\bibinfo {author} {\bibfnamefont {F.}~\bibnamefont
  {Yang}}, \bibinfo {author} {\bibfnamefont {P.~A.}\ \bibnamefont {Huidobro}},
  \ and\ \bibinfo {author} {\bibfnamefont {J.~B.}\ \bibnamefont {Pendry}},\
  }\href@noop {} {\bibfield  {journal} {\bibinfo  {journal} {Physical Review
  B}\ }\textbf {\bibinfo {volume} {98}},\ \bibinfo {pages} {125409} (\bibinfo
  {year} {2018})}\BibitemShut {NoStop}%
\bibitem [{\citenamefont {Khurgin}(2015)}]{khurgin2015deal}%
  \BibitemOpen
  \bibfield  {author} {\bibinfo {author} {\bibfnamefont {J.~B.}\ \bibnamefont
  {Khurgin}},\ }\href@noop {} {\bibfield  {journal} {\bibinfo  {journal}
  {Nature Nanotechnology}\ }\textbf {\bibinfo {volume} {10}},\ \bibinfo {pages}
  {2} (\bibinfo {year} {2015})}\BibitemShut {NoStop}%
\bibitem [{\citenamefont {Fukui}\ \emph {et~al.}(1979)\citenamefont {Fukui},
  \citenamefont {So},\ and\ \citenamefont {Normandin}}]{fukui1979lifetimes}%
  \BibitemOpen
  \bibfield  {author} {\bibinfo {author} {\bibfnamefont {M.}~\bibnamefont
  {Fukui}}, \bibinfo {author} {\bibfnamefont {V.}~\bibnamefont {So}}, \ and\
  \bibinfo {author} {\bibfnamefont {R.}~\bibnamefont {Normandin}},\ }\href@noop
  {} {\bibfield  {journal} {\bibinfo  {journal} {Physica Status Solidi (b)}\
  }\textbf {\bibinfo {volume} {91}},\ \bibinfo {pages} {K61} (\bibinfo {year}
  {1979})}\BibitemShut {NoStop}%
\bibitem [{\citenamefont {West}\ \emph {et~al.}(2010)\citenamefont {West},
  \citenamefont {Ishii}, \citenamefont {Naik}, \citenamefont {Emani},
  \citenamefont {Shalaev},\ and\ \citenamefont
  {Boltasseva}}]{west2010searching}%
  \BibitemOpen
  \bibfield  {author} {\bibinfo {author} {\bibfnamefont {P.~R.}\ \bibnamefont
  {West}}, \bibinfo {author} {\bibfnamefont {S.}~\bibnamefont {Ishii}},
  \bibinfo {author} {\bibfnamefont {G.~V.}\ \bibnamefont {Naik}}, \bibinfo
  {author} {\bibfnamefont {N.~K.}\ \bibnamefont {Emani}}, \bibinfo {author}
  {\bibfnamefont {V.~M.}\ \bibnamefont {Shalaev}}, \ and\ \bibinfo {author}
  {\bibfnamefont {A.}~\bibnamefont {Boltasseva}},\ }\href@noop {} {\bibfield
  {journal} {\bibinfo  {journal} {Laser \& Photonics Reviews}\ }\textbf
  {\bibinfo {volume} {4}},\ \bibinfo {pages} {795} (\bibinfo {year}
  {2010})}\BibitemShut {NoStop}%
\bibitem [{\citenamefont {Naik}\ \emph {et~al.}(2013)\citenamefont {Naik},
  \citenamefont {Shalaev},\ and\ \citenamefont
  {Boltasseva}}]{naik2013alternative}%
  \BibitemOpen
  \bibfield  {author} {\bibinfo {author} {\bibfnamefont {G.~V.}\ \bibnamefont
  {Naik}}, \bibinfo {author} {\bibfnamefont {V.~M.}\ \bibnamefont {Shalaev}}, \
  and\ \bibinfo {author} {\bibfnamefont {A.}~\bibnamefont {Boltasseva}},\
  }\href@noop {} {\bibfield  {journal} {\bibinfo  {journal} {Advanced
  Materials}\ }\textbf {\bibinfo {volume} {25}},\ \bibinfo {pages} {3264}
  (\bibinfo {year} {2013})}\BibitemShut {NoStop}%
\end{thebibliography}%

\end{document}